\pdfoutput=1
\documentclass[%
 reprint,
superscriptaddress,
 amsmath,amssymb,
 aps,
prb,
floatfix,
]{revtex4-1}

\usepackage{graphicx}
\usepackage{dcolumn}
\usepackage{bm}
\usepackage[colorlinks=true,allcolors=blue]{hyperref}
\usepackage{array,booktabs,tabularx}
\usepackage[caption=false]{subfig}
\usepackage[USenglish]{babel}
\usepackage{braket}

\newcommand{\abinitio}{\emph{ab initio}}
\newcommand{\beq}{\begin{equation}}
\newcommand{\eeq}{\end{equation}}
\newcommand{\beqn}{\begin{eqnarray}}
\newcommand{\eeqn}{\end{eqnarray}}

\begin{document}
\selectlanguage{USenglish}
\preprint{APS/123-QED}

\title{Anharmonic effects in atomic hydrogen: superconductivity and lattice dynamical stability}

\author{Miguel Borinaga}
\affiliation{Centro de F\'isica de Materiales CFM, CSIC-UPV/EHU, Paseo Manuel de
             Lardizabal 5, 20018 Donostia/San Sebasti\'an, Basque Country, Spain}
\affiliation{Donostia International Physics Center
             (DIPC), Manuel Lardizabal pasealekua 4, 20018 Donostia/San
             Sebasti\'an, Basque Country, Spain}
\author{Ion Errea}
\affiliation{Donostia International Physics Center
             (DIPC), Manuel Lardizabal pasealekua 4, 20018 Donostia/San
             Sebasti\'an, Basque Country, Spain}
\affiliation{Fisika Aplikatua 1 Saila, EUITI Bilbao,
             University of the Basque Country (UPV/EHU), Rafael Moreno ``Pitxitxi'' Pasealekua 3, 48013 Bilbao,
             Basque Country, Spain}
\author{Matteo Calandra}
\affiliation{IMPMC, UMR CNRS 7590, Sorbonne
Universit\'es - UPMC Univ. Paris 06, MNHN, IRD, 4 Place Jussieu,
F-75005 Paris, France}
\author{Francesco Mauri}
\affiliation{IMPMC, UMR CNRS 7590, Sorbonne
Universit\'es - UPMC Univ. Paris 06, MNHN, IRD, 4 Place Jussieu,
F-75005 Paris, France}
\affiliation{Dipartimento di Fisica, Universit\`a di Roma La Sapienza, Piazzale Aldo Moro 5, I-00185 Roma, Italy}
\author{Aitor Bergara}
\affiliation{Centro de F\'isica de Materiales CFM, CSIC-UPV/EHU, Paseo Manuel de
             Lardizabal 5, 20018 Donostia/San Sebasti\'an, Basque Country, Spain}
\affiliation{Donostia International Physics Center
             (DIPC), Manuel Lardizabal pasealekua 4, 20018 Donostia/San
             Sebasti\'an, Basque Country, Spain}
\affiliation{Departamento de F\'isica de la Materia Condensada,  University of the Basque Country (UPV/EHU), 48080 Bilbao, 
             Basque Country, Spain}



\date{\today}

\begin{abstract}

We present first-principles calculations of metallic atomic hydrogen in the 400-600 GPa pressure range in a tetragonal 
structure with space group $I4_1/amd$,
which is predicted to be its first atomic phase. 
Our calculations show a band structure close to the free-electron-like limit due to the high 
electronic kinetic energy induced by pressure. 
Bands are properly described even in the independent electron approximation 
fully neglecting the electron-electron interaction. 
Linear-response harmonic calculations show a dynamically stable phonon spectrum with marked Kohn anomalies. 
Even if the electron-electron interaction has a minor role in the electronic bands, 
the inclusion of electronic exchange and correlation in the density response
is essential to obtain a dynamically stable structure. 
Anharmonic effects, which are calculated within the stochastic self-consistent harmonic approximation, 
harden high-energy optical modes and soften transverse acoustic modes up to a 20\% in energy.
Despite a large impact of anharmonicity has been predicted in several high-pressure hydrides,
here the superconducting critical temperature is barely affected by anharmonicity, as it is 
lowered from its harmonic 318 K value only to 300 K at 500 GPa. We atribute the small impact of anharmoncity on 
superconductivity to the absence of softened optical modes and the fairly uniform distribution of the electron-phonon 
coupling among the vibrational modes.  

\end{abstract}

\maketitle


\section{Introduction}\label{introduction}

The recent measurement of a superconducting critical temperature ($T_c$) of 203 K
in the sulfur hydrogen system\cite{Drozdov2015}, a temperature
reachable on Earth's surface, is a major breakthrough
in the field of superconductivity. It ultimately validates
Ashcroft's idea that hydrogen and hydrogen-dominant metallic compounds
can be high-temperature
superconductors~\cite{PhysRevLett.21.1748,PhysRevLett.92.187002}.
This measurement offers new hopes to find sooner than later 
room-temperature superconductivity in other hydrogen-rich compounds or
hydrogen itself. Indeed, since the advent of modern \abinitio{} calculations
for the electron-phonon interaction, many theoretical calculations
have predicted astonishingly high $T_c$ values for hydrogen at megabar
pressures, both in molecular and atomic 
phases, the largest $T_c$ values predicted for any 
compound up to date\cite{cudazzo:257001,2007SSCom.141..610Z,PhysRevB.81.134505,PhysRevB.81.134506,
PhysRevB.84.144515,Yan20111264,Maksimov2001569}.

Thus far, five solid phases of hydrogen have been determined, all of them  molecular and insulating. However, a sixth metallic one
has been recently claimed to be found\cite{Eremets2016}. 
Phase I is thought to be a molecular solid of quantum rotors on a hexagonal 
close packed lattice\cite{Nat.383.702}. It is found in a wide pressure and temperature range 
up to the melting curve, which has a maximum of 
$\sim$1000 K around 65 GPa\cite{PhysRevLett.100.155701,natureliquidh}. Phase
II appears between approximately 50-150 GPa and only below 150 K, 
temperature at which it transforms back to phase I\cite{RevModPhys.66.671}. 
Hydrogen adopts phase III above 150 GPa up to at least 
360 GPa\cite{PhysRevLett.108.146402,Goncharov04122001}. Recent experiments have determined that this phase 
transforms to phase IV at around 200 K in the 240-325 GPa pressure range 
\cite{Eremets2011,PhysRevLett.108.125501,PhysRevLett.110.217402},
which might melt close to room temperature\cite{Howie2015}. Transition from phase IV to V occurs at 325 GPa, 
the last one existing probably up to the dissociation pressure\cite{Dalladay-Simpson2016}. Finally, a new  
work claims to have found phase VI by cooling down phase V below 200-220 K. This new phase could be metallic, 
but further work is expected to confirm this result\cite{Eremets2016}. 
In any case, and even though the goal seems closer than ever, the quest for metallic hydrogen continues.
Extrapolation of optical experiments\cite{optical} predicted a metallization pressure of around $450$ GPa, which was expected to 
occur due to the overlap between the valence and conduction bands in the molecular state before molecular dissociation\cite{pickard:473,PhysRevB.85.214114}.
However, works in Ref.~\onlinecite{Dalladay-Simpson2016} and~\onlinecite{ Eremets2016} suggest metallization might occur together 
with molecular dissociation. In any case, state-of-the-art static diamond anvil cell techniques\cite{Dubrovinsky2015} will allow promising future experiments that are expected
to shine light on this long standing quest.

\begin{figure}[t]
\subfloat[][]{\includegraphics[width=0.5\linewidth]{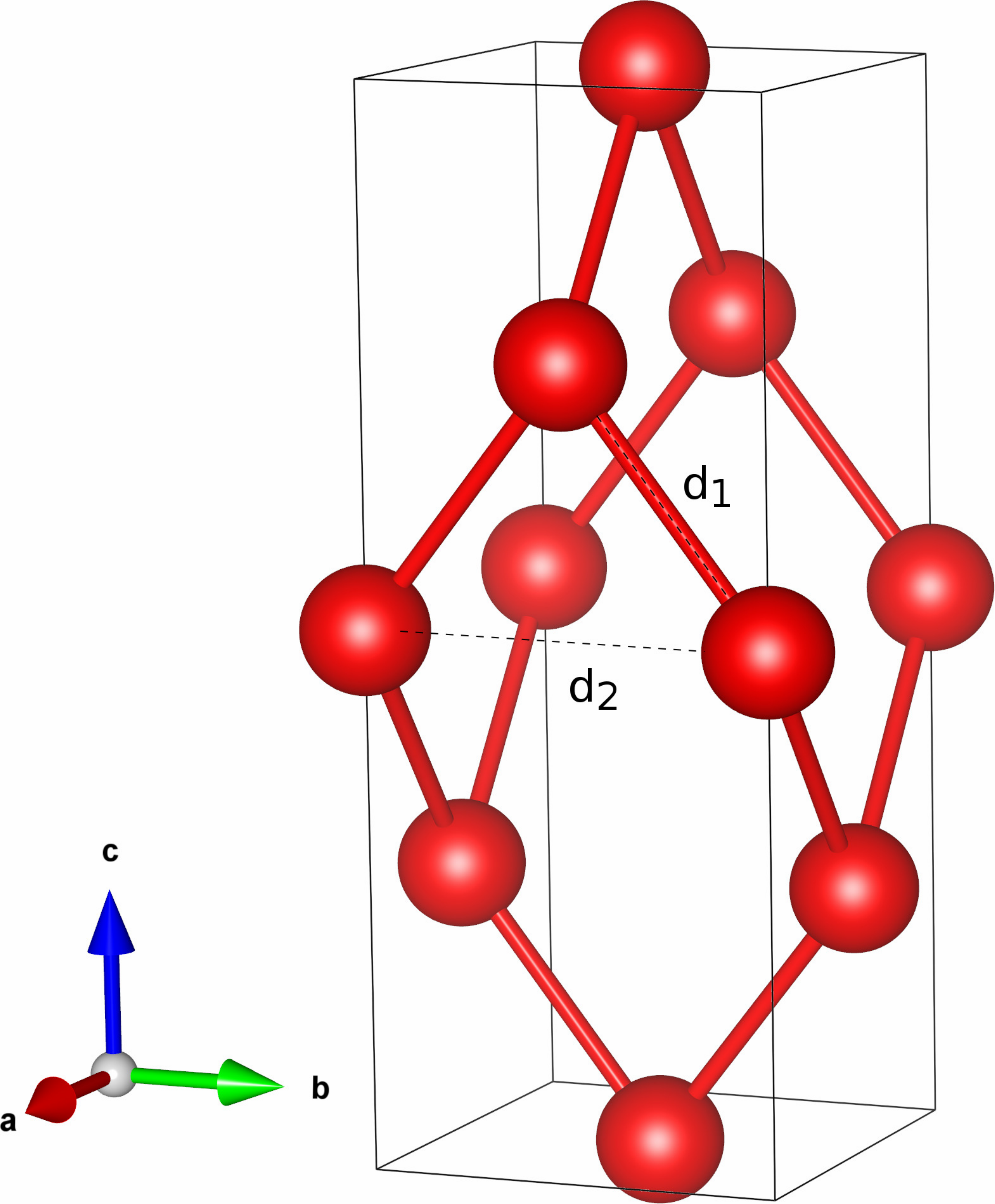}}\hspace{0.1\linewidth}
\subfloat[][]{
\begin{tabular}[b]{|c|c|}
    \hline
    $a=b$ & 1.21 \AA \\
    $c$ & 3.08 \AA \\
    \hline
    $d_1$ & 0.98 \AA \\
    $d_2$ & 1.21 \AA \\
    \hline
  \end{tabular}
}
\caption{\label{fig:structure} (a) Conventional unit cell of $I4_1/amd$ hydrogen. (b) $a=b$ and $c$ lattice parameters as well as the two different interatomic distances $d_1$ and $d_2$ of $I4_1/amd$ hydrogen at 500 GPa.}
\end{figure}

According to recent quantum Monte Carlo calculations including anharmonicity for the 
zero-point energy\cite{PhysRevLett.112.165501}, above 374 GPa hydrogen should undergo 
a transition to a metallic and atomic tetragonal $I4_1/amd$ phase (shown in Fig. \ref{fig:structure}), which had been predicted before\cite{PhysRevLett.106.165302}.
It seems the inclusion of anharmonicity is important to describe the boundaries of
phases I, II, III, and IV\cite{Drummond2015}, as well as to estimate the vibron 
energies in phases III and IV\cite{PhysRevB.87.174110,PhysRevLett.110.025903}.
Nevertheless, even if anharmonicity has a huge impact on the superconducting 
properties of many hydrides\cite{PhysRevLett.114.157004,errea_inverse_isotope,PhysRevB.89.064302,
doi:10.1080/08957959.2010.520209,PhysRevB.82.104504},
the potential impact of anharmonicity in the large $T_c$ values predicted for 
hydrogen\cite{cudazzo:257001,2007SSCom.141..610Z,PhysRevB.81.134505,PhysRevB.81.134506,
PhysRevB.84.144515,Yan20111264,Maksimov2001569} remains unexplored.
Considering that in aluminum\cite{doi:10.1080/08957959.2010.520209,PhysRevB.82.104504}, palladium\cite{errea_inverse_isotope}, 
and platinum hydrides\cite{PhysRevB.89.064302}, as well as in the record superconductor H$_3$S\cite{PhysRevLett.114.157004}
anharmonicity suppresses the electron-phonon couping as it tends to harden optical H-character  
phonon modes, one might expect that anharmonicity might strongly impact superconductivity 
in hydrogen too.

In this work we present an \abinitio{} analysis based on density-functional theory (DFT) 
of electronic and vibrational properties of 
$I4_1/amd$ hydrogen in the 400-600 GPa pressure range. Our goal is to elucidate what the impact of anharmonicity
is in the possible first structure of atomic hydrogen. Our results show a close to free-electron-like behavior, 
as the DFT band structure is perfectly described even if electron-electron interaction is fully neglected. 
The harmonic phonons calculated within linear-response 
density-functional perturbation theory (DFPT)\cite{RevModPhys.73.515} show the system is 
dynamically stable even if marked Kohn anomalies are present.
Despite the minor role of electron-electron interaction in the electronic states,
electronic exchange becomes of vital importance to ensure the dynamical stability of the system. 
We have included anharmonic effects using the 
stochastic self-consistent harmonic approximation (SSCHA)\cite{errea_inverse_isotope,PhysRevB.89.064302}
and observed the phonon spectrum is significantly modified. However, in contrast to 
many superconducting hydrides\cite{PhysRevLett.114.157004,errea_inverse_isotope,PhysRevB.89.064302,
doi:10.1080/08957959.2010.520209,PhysRevB.82.104504},
the superconducting $T_c$ obtained within the harmonic approximation is barely modified by anharmonicity,
it is only suppressed by a 6\%. 

The paper is organized as follows. In Sec. \ref{theory} we overview 
the computational details and methods used throughout. In Sec. \ref{results} we 
present and analyze the obtained results. Finally, we present a summary of our conclusions
in Sec. \ref{conclusions}.

\section{Computational details}\label{theory}

We performed our DFT calculations within the Perdew-Burke-Ernzerhof 
parametrization of the generalized-gradient approximation\cite{pbe}.
The electron-proton interaction was considered making use of an 
ultrasoft pseudopotential as implemented in {\sc Quantum ESPRESSO}~\cite{0953-8984-21-39-395502}.
Due to the large kinetic energy of the electrons, a proper convergence of 
the electronic properties and phonon frequencies
required a dense $80\times80\times80$ \textbf{k}-mesh and $0.05$ Ry Hermite-Gaussian
electronic smearing for the electronic integrations in the first Brillouin zone (BZ). 
An energy cutoff of $100$ Ry was necessary for expanding
the wave-functions in the plane-wave basis. 

Phonon frequencies were calculated within DFPT as implemented in {\sc Quantum ESPRESSO}~\cite{0953-8984-21-39-395502}
in a $6\times6\times6$ \textbf{q}-point grid in the BZ. Fourier interpolation was used to 
obtain the phonon spectra along high-symmetry lines. We find convenient to split the
calculated dynamical matrices at a given wave-vector as $D(\mathbf{q})=D_{p}(\mathbf{q})+D_{e}(\mathbf{q})$.
$D_p$ represents the contribution of the proton-proton Coulomb interaction to the dynamical matrix, which 
can be estimated analytically with an Ewald summation. $D_e$ contains the effect of the electronic
response to the proton motion in the dynamical matrix. Besides DFPT, we have also 
estimated $D_e$ making use of the free-electron Lindhard
response function within the Random Phase Approximation (RPA) as in Ref. \onlinecite{Rousseau2010}.
In the latter approach the dynamical matrix is analytical at any ${\bf q}$ so that we have not
restricted the calculations to the $6\times6\times6$ grid. 

Electron-phonon martix elements were also calculated within DFPT in a $6\times6\times6$ ${\bf q}$-point 
grid. Converging the double Dirac delta
in the equation for the phonon linewidth required a $100\times100\times100$ denser \textbf{k}-point mesh.
The superconducting $T_c$ was calculated solving isotropic Migdal-Eliashberg 
equations\cite{sov.phys.jetp.7.996,sov.phys.jetp.11.696}, considering that
for large electron-phonon coupling constants McMillan's equation 
underestimates $T_c$\cite{PhysRevB.12.905}. 

We use the SSCHA\cite{errea_inverse_isotope,PhysRevB.89.064302} to calculate the anharmonic renormalization
of the phonon spectrum. The SSCHA is a variational method 
in which the vibrational free energy is minimized with respect to a trial harmonic 
density matrix. This minimization process requires the 
calculation of forces acting on atoms in supercells for different configurations 
created with the trial density matrix. 
These forces were calculated in a $3\times3\times3$ supercell 
making use of DFT with the same parameters as the DFPT phonon calculations.
This yielded phonon frequencies in a commensurate $3\times3\times3$ ${\bf q}$-point grid.
The difference between the harmonic and anharmonic dynamical matrices was interpolated
to the finer $6\times6\times6$ grid. 
The anharmonic correction to the superconducting $T_c$ was calculated
combining the SSCHA dynamical matrices with the calculated electron-phonon deformation
potential within DFPT as explained in Ref. \onlinecite{errea_inverse_isotope}.

\section{Results and discussion}\label{results}

\subsection{Electronic structure}

\begin{figure*}[t]
\subfloat[(a)][]{\includegraphics[width=0.48\linewidth]{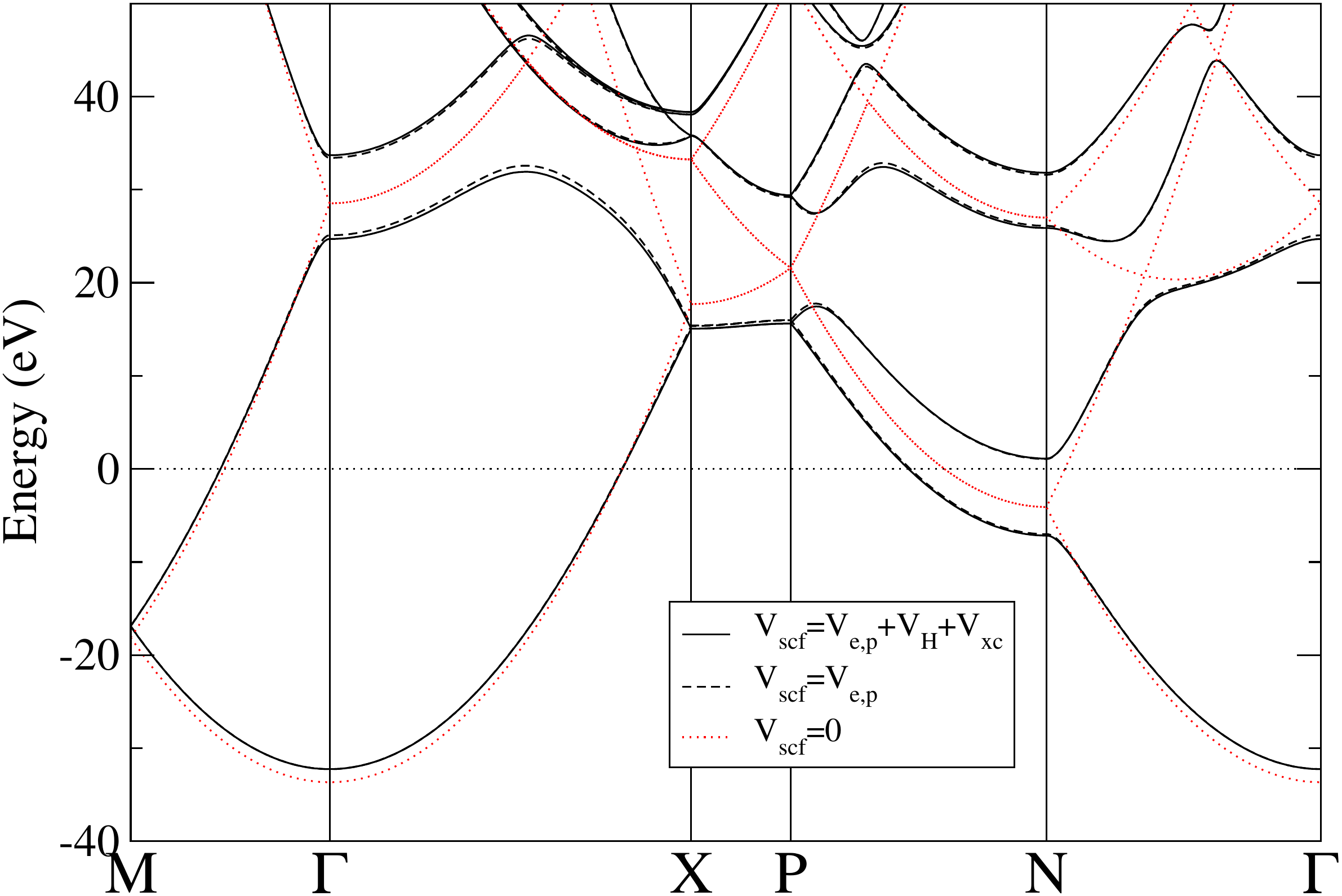}\label{fig:bands}}\hfill
\subfloat[(b)][]{\includegraphics[width=0.35\linewidth]{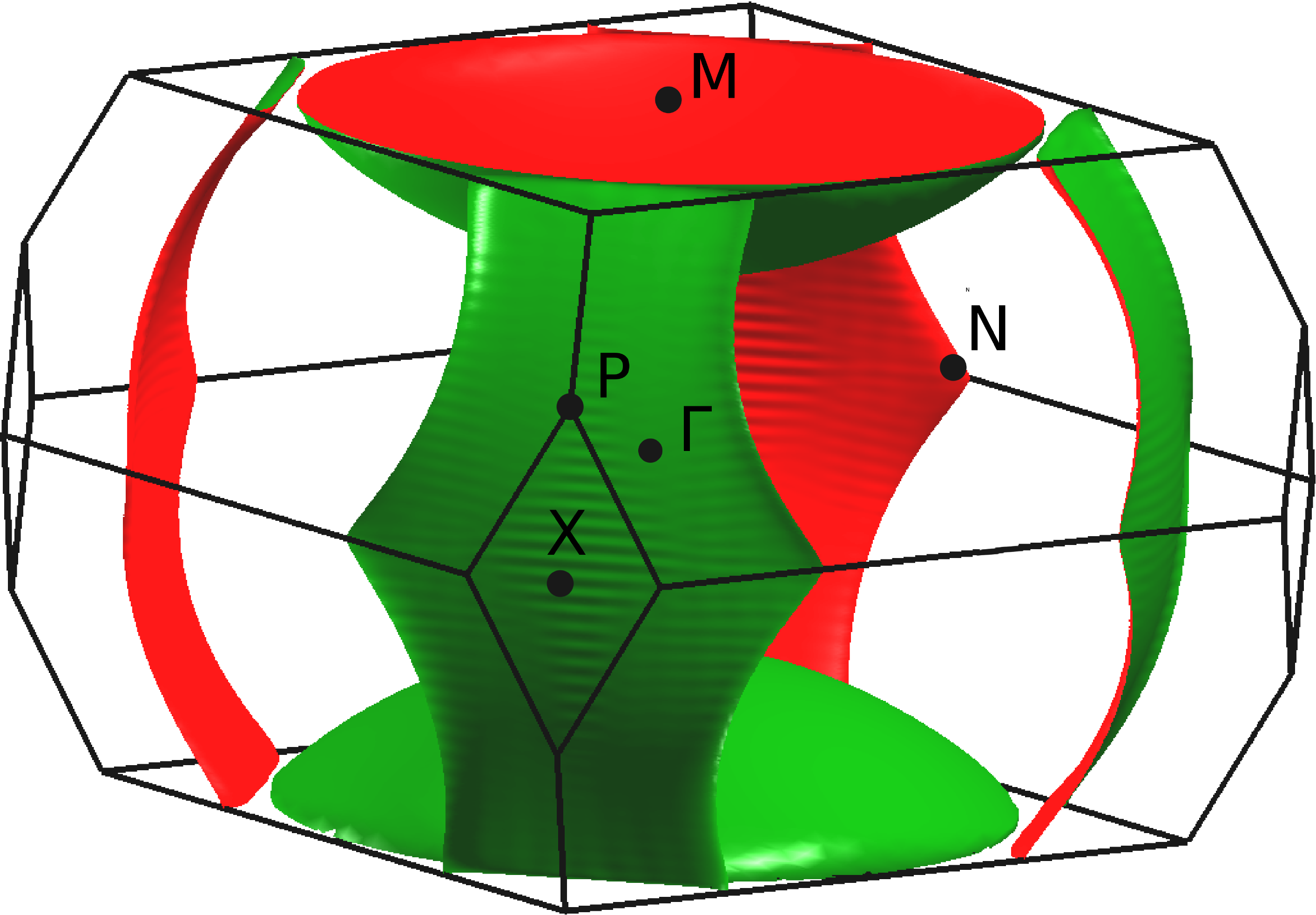}\label{fig:fermisurface}}\hfill~
\caption{(a) Electronic band structure of $I4_1/amd$ hydrogen at $500$ GPa
calculated within DFT, where the self-consistent potential is
$V_{scf} = V_{e,p}+V_H+V_{xc}$. 
Bands obtained  within the free-electron approach ($V_{scf}=0$) 
and the independent-electron approach ($V_{scf}=V_{e,p}$)
are also shown.
The origin of energy (black dotted line) corresponds to the Fermi level. 
(b) Fermi surface of $I4_1/amd$ hydrogen at $500$ GPa within DFT.
The BZ and its high-symmetry points are shown.}
\end{figure*}

Hydrogen, containing a single proton and an electron, is supposed to leave its common 
molecular nature under a sufficient high pressure to form a metal similar to the alkalies.
In the high pressure limit the dominant electronic kinetic energy would be 
responsible for ending up with a free-electron-like metal. However, in the analyzed pressure range we are still close to
the dissociation/metallization pressure\cite{optical} and the problem is not as simple as one could 
imagine \textit{a priori}.

Fig. \ref{fig:bands} shows the electronic band structure of $I4_1/amd$ hydrogen at 500 GPa. 
The bands present a huge dispersion, associated to the dominating kinetic term in 
the energies of the electronic states. 
The calculated band structure is not far 
from the free-electron approximation, the main difference being the band gaps opened at the border of the BZ 
and whenever band crossing occurs due to the interaction of the electrons with the proton lattice. 
It is interesting to point out that 
if we consider the independent electron approximation, just keeping the $V_{e,p}$ term 
that gives the electron-proton interaction
in the $V_{scf}$ self-consistent potential and neglecting the $V_H$ Hartree
and $V_{xc}$ electron-electron interactions 
(see \hyperref[appendix]{Appendix} for details), the 
resulting bands match almost perfectly with the DFT ones. Therefore, we can 
conclude that main differences with the free-electron approximation are due to the 
large proton-electron interaction, and that the interaction between 
the electrons is not giving any significant contribution to the band structure.
Fig. \ref{fig:fermisurface} shows the Fermi surface, which is quite spherical. However, the sphere shows some open areas 
around the high symmetry point N, where it touches the BZ boundary and a band gap is opened.

\subsection{Vibrational properties}\label{vibrations}

\begin{figure}[t]
\includegraphics[width=1.0\linewidth]{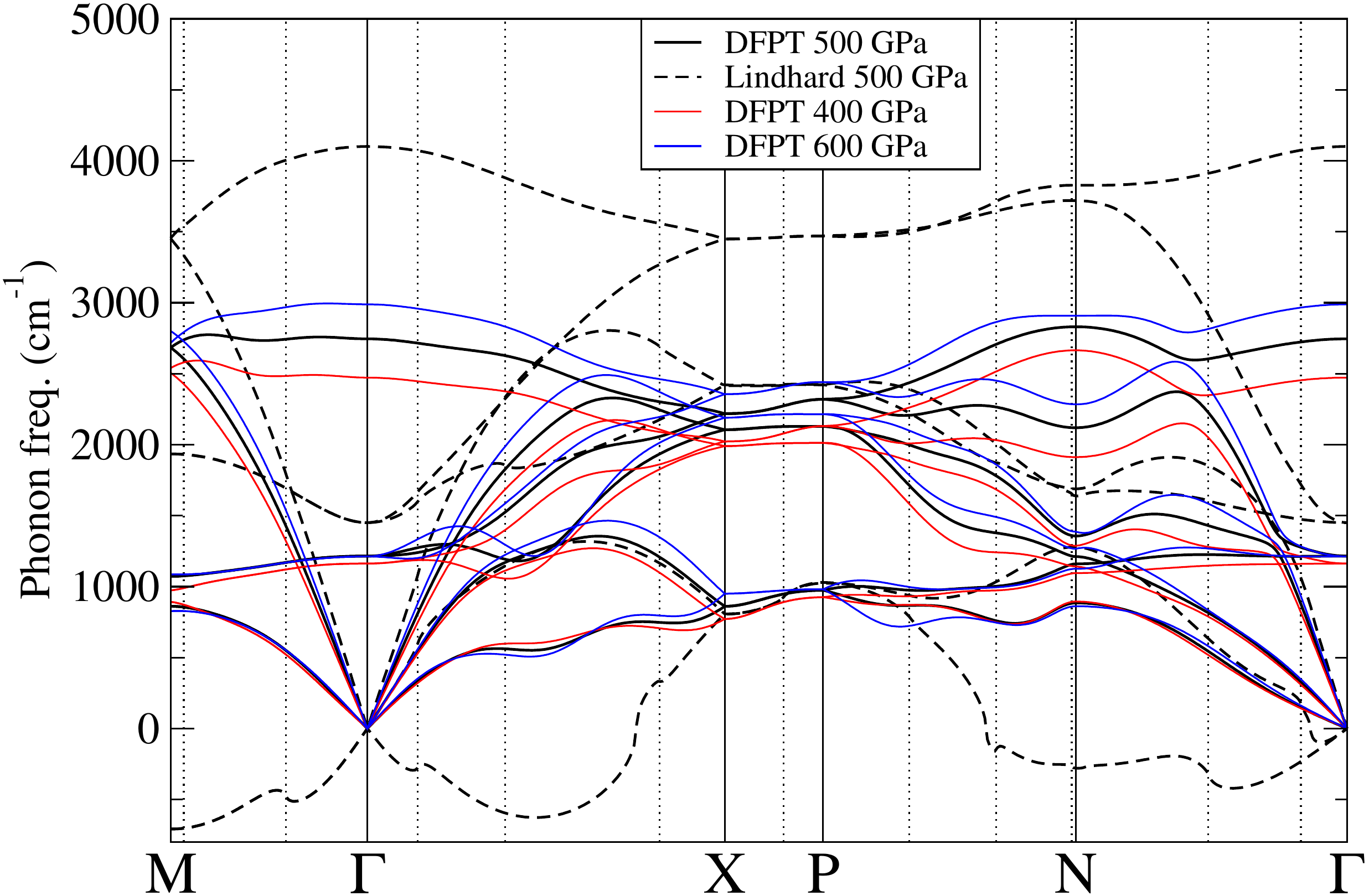}
\caption{\label{fig:phonons.lindhard} Phonon spectra of $I4_1/amd$ hydrogen calculated within DFPT at several pressures. At 500 GPa phonons calculated within the Lindhard RPA formulation are also shown. 
Dotted vertical black lines indicate \textbf{q}-points satisfying
$|\mathbf{q+G}|=2k_f$, where Kohn anomalies are expected to appear.}
\end{figure}

In Fig. \ref{fig:phonons.lindhard} we show the calculated phonon dispersion in tetragonal $I4_1/amd$ hydrogen 
at 500 GPa within DFPT. We also show the phonons at 400 and 600 GPa, but due to the minor qualitative changes we will focus just
in the 500 GPa spectrum, as the analysis should be valid for all the pressure range. The system clearly is dynamically stable, but 
there are some branches with strong Kohn anomalies\cite{PhysRevLett.90.035501}, specially the low energy transverse acoustic branch.
Indeed, as shown in Fig. \ref{fig:phonons.lindhard}, \textbf{q}-points at which the anomalies appear
coincide with $|\mathbf{q+G}|=2k_f$, where $k_f$ is the Fermi wave-vector and $\mathbf{G}$ the reciprocal lattice vector
that brings $\mathbf{q}$ back into the BZ.
Considering the validity of the free-electron-like approximation to describe the 
electronic band structure, we have calculated the phonon dispersion within this approximation, 
assuming that the $D_e$ contribution to the dynamical matrix can be 
calculated with the Lindhard response function
at the RPA level as in Ref. \onlinecite{Rousseau2010}, therefore, 
neglecting correlation and exchange in the electronic response. 
In this free-electron limit we can  obtain the spectra along high-symmetry lines without any Fourier interpolation,
evidencing the presence of the Kohn anomalies at the $|\mathbf{q+G}|=2k_f$ points, and confirming that
the kinks present in the DFPT result are Kohn anomalies. 

\begin{figure}[t]
\includegraphics[width=1.0\linewidth]{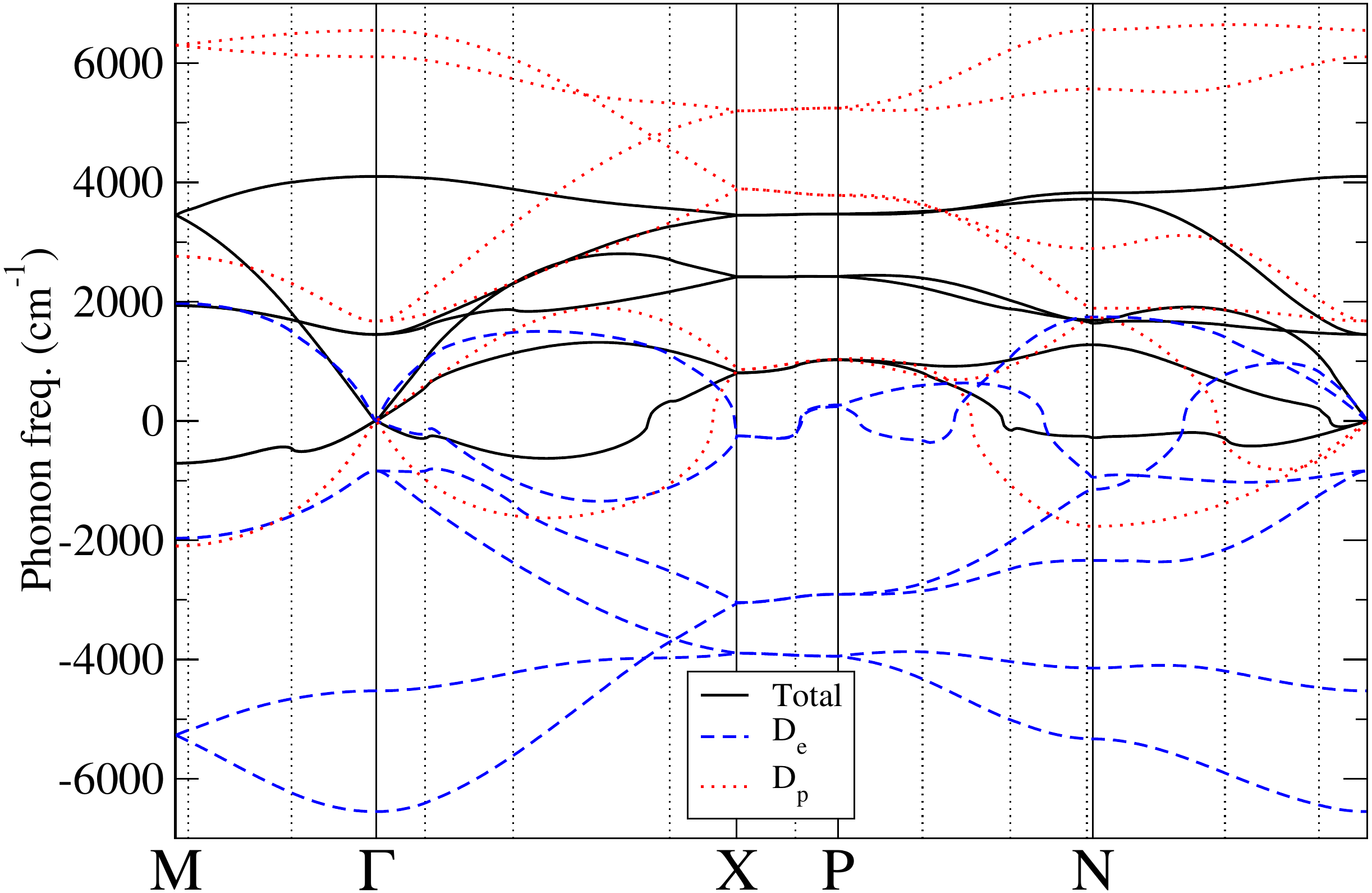}\hspace{0.1cm}
\caption{\label{fig:contributions} Lindhard RPA phonons of $I4_1/amd$ hydrogen at $500$ GPa, decomposed 
into the contributions coming from the $D_p$ 
and $D_e$ terms. Dotted vertical lines indicate \textbf{q}-points satisfying
$|\mathbf{q+G}|=2k_f$, where Kohn anomalies appear.}
\end{figure}

Nevertheless, the Lindhard RPA  spectrum completely differs from the \abinitio{} calculations. 
The intensity of the Kohn anomalies is much stronger and the transverse acoustic modes
become unstable with imaginary frequencies. Therefore, 
even though the electronic band structure could be understood within 
the free-electron-like approximation, phonons seem to be far from this picture, contrary
to the case of sodium\cite{Rousseau2010,springerlink:10.1140/epjb/e2011-10972-9}.  
This fact also questions the stability of the $I4_1/amd$ tetragonal
phase in the ultimate high-pressure limit where the electrons are expected to be free.  
If we split the dynamical matrix in the $D_p$ and $D_e$ terms, we note that each of them 
scales differently with the average inter-electronic distance parameter $r_s$:
$D_p$ always scales as $r_s^{-3}$, while, in the Lindhard RPA, $D_e$ scales as $r_s^{-2}(r_s+C)$, where $C$ is always positive 
and of the order of unity. Therefore, in the very large pressure limit with small $r_s$,
the $D_p$ contribution is expected to dominate over the electronic contribution. 
In Fig. \ref{fig:contributions} we present the dispersion of the root of the eigenvalues of 
each contribution separately. This represents the phonon spectra that would be
obtained from each contribution independently. In our case, phonons associated to $D_p$ are already unstable and the contribution 
from $D_e$ is not enough to stabilize them. This is the reason why the Lindhard RPA phonons have imaginary
frequencies. As in the high-pressure limit $D_p$ will dominate over $D_e$, the tetragonal $I4_1/amd$ 
will not become stable at very large pressure, but more symmetric and compact structures with positive eigenvalues
of $D_p$ (as fcc or bcc) will be favored.  

In order to better understand the discrepancy between the \abinitio{} DFPT 
and the Lindhard RPA spectra, we have made several calculations based on linear-response theory
trying to disentangle the different contributions to the final phonon spectra.
The calculation of $D_e$ 
requires not only the knowledge of the electronic density 
$n$ but also its linear change $\delta n$ with respect to 
ionic displacementsi~\cite{RevModPhys.73.515}.
This last term can be calculated from the interacting 
electronic density-response function $\chi$, which requires first
the knowledge of the non-interacting response $\chi_0$, and second to solve an integral Dyson-like
equation where the linear change of the self-consistent potential $\delta V_{scf}$ takes part
(see \hyperref[appendix]{Appendix}). 
Although it requires a cumbersome sum over excited state, 
$\chi_0$ can be calculated directly from the eigenvalues and eigenstates
obtained with the unperturbed $V_{scf}$ self-consistent potential.
Combining different approximations in $V_{scf}$ to calculate $\chi_0$
and $\delta V_{scf}$ to calculate $\chi$, the relevant contributions 
to the dynamical matrices can be unveiled.
For instance, in the Lindhard RPA limit $\chi_0$ is built 
from free-electron bands ($V_{scf}=0$) and exchange-correlation is neglected to build $\chi$
($\delta V_{scf} = \delta V_{e,p} + \delta V_H$), limit in which
$D_e$ is reduced to a simple analytical expression~\cite{Rousseau2010}. 
Here, we avoid the explicit calculation of $\chi_0$ and $\chi$ by calculating
$\delta n$ making use of DFPT. In this case, the Sternheimer equation 
is solved self-consistently neglecting 
different terms in $V_{scf}$ and $\delta V_{scf}$, which is equivalent to
making different approaches to $\chi_0$ and $\chi$ respectively.
In those cases when a different scheme is adopted for $V_{scf}$ and $\delta V_{scf}$
translational invariance is not satisfied anymore. We overcome this
difficulty imposing the acoustic sum rule (ASR) \emph{a posteriori}
(see \hyperref[appendix]{Appendix} for technical details). 

\begin{figure}[t]
\includegraphics[width=1.0\linewidth]{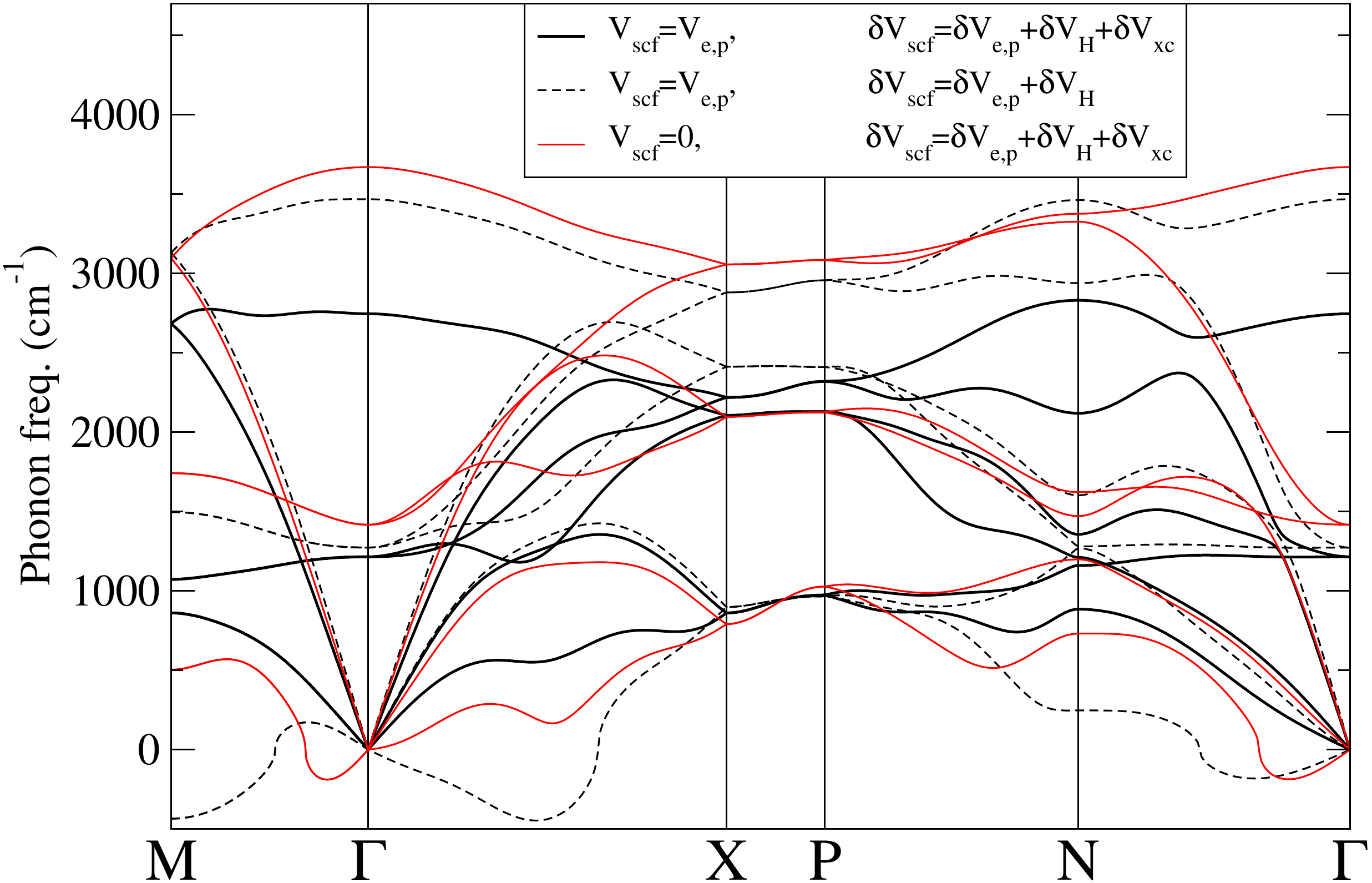}\hspace{0.1cm}
\caption{\label{fig:phonons.all} Phonon spectra of $I4_1/amd$ hydrogen at $500$ GPa within
different approximations for the unperturbed self-consitent potential $V_{scf}$ and its 
linear perturbation $\delta V_{scf}$. Calculations are performed  within the DFPT formalism. 
The ASR has been imposed \textit{a posteriori}. 
}
\end{figure}

In Fig. \ref{fig:phonons.all} we show three phonon spectra with different combination 
of neglected terms in $V_{scf}$ and $\delta V_{scf}$. First, we show the spectrum obtained 
by neglecting any
electron-electron interaction in the unperturbed Hamiltonian ($V_{scf}=V_{e,p}$), 
but keeping all the terms in its linear perturbation $\delta V_{scf}=\delta V_{e,p}+\delta V_{H}+\delta V_{xc}$.
We obtain exactly the same result as in the previously calculated full 
DFPT phonons (see Fig. \ref{spectraappendix2} in \hyperref[appendix]{Appendix} for a comparison), 
remarking the insignificant role of electron-electron interaction 
in the electronic bands and, consequently, in $\chi_0$. Second, we show the phonons calculated neglecting electron-electron interaction in $V_{scf}$ again, but this time neglecting the exchange and correlation term in $\delta V_{scf}$. 
This is equivalent to calculating $\chi_0$ as in the previous case, but calculating $\chi$ within the
RPA. We can see that electronic exchange and correlation is crucial to properly account for 
the electronic response in this system as the RPA fails 
dramatically predicting that the system is dynamically unstable. 
Finally, we show the dispersion obtained 
with free electrons ($V_{scf}=0$) but using the full linear perturbation going beyond the RPA. 
Comparing the free-electron RPA calculation in Fig. \ref{fig:phonons.lindhard}
with the one including exchange-correlation effects for the response in Fig. \ref{fig:phonons.all},
it can be confirmed again that including exchange and correlation in $\delta V_{scf}$ 
(and therefore in $\chi$) is determinant for obtaining dynamically stable phonons.  
However, due to the strong electron-proton interaction, 
including $V_{e,p}$ in the self-consistent potential is necessary to obtain
good quantitative results. 

Analyzing these different calculations, we conclude that, 
despite the electronic kinetic energy dominates and the electron-electron
interaction plays a negligible role in the band structure, 
going beyond the RPA including exchange-correlation effects in the 
calculation of the electronic response is crucial. 
Indeed, exchange-correlation effects in the response to the proton motion 
make tetragonal $I4_1/amd$ atomic hydrogen dynamically stable. This calamitous failure of the RPA 
is related to the high electronic density around the protons, 
which is underestimated by the Hartree term by up to a $9\%$. 
We have checked that similar phonons are obtained 
for different approximations of the exchange-correlation potential\cite{PhysRevB.37.785,ldalinear}. 
This is not the case in the molecular case at lower pressures, 
where correlation clearly has an effect in the energies of molecular vibrons\cite{PhysRevB.88.014115}.

Even if exchange-correlation effects guarantee the stability of the tetragonal
$I4_1/amd$ atomic structure of hydrogen, this conclusion is drawn exclusively
at the harmonic level.
In Fig. \ref{fig:phonons.anharmonic} we compare the harmonic DFPT phonon spectra with the
anharmonic spectra obtained within the SSCHA.
Anharmonicity is quite strong, as phonon frequencies are affected up to 
a $20\%$ in the transverse acoustic branch in the $\Gamma$-N path. 
Transverse acoustic branches are in general specially vulnerable
as their energies are considerably lowered. 
Consequently, one could expect anharmonic effects to increase the zero-point displacement of the atoms, 
bringing atomic hydrogen closer to quantum melting. According to our calculations, however,
$I4_1/amd$ hydrogen is not melted if we hold to the quantum limit of the Lindemman criterion,
which states quantum melting occurs when the Root Mean Square (RMS) displacements of atoms are around $27-30\%$
of the interatomic distances \cite{PhysRevB.55.5767}. While in the harmonic approach 
we obtain RMS displacements of a $20.3\%$ and a $16.4\%$ relative to the two different 
interatomic distances $d_1$ and $d_2$ of the crystal, 
these values are only slightly raised by anharmonicity to a $20.7\%$ and a $16.6\%$, 
respectively. The reason of this rather small change is a big part of the high 
energy modes are enhanced by anharmonicity, 
compensating the softening of the transverse acoustic branches. 

As most of the phases have been found and characterized by Raman and infrared 
spectroscopy experiments, we show the impact of anharmonicity in the optical modes at the 
$\Gamma$ point in 
Table \ref{table:ramanir}. 
The structure has two Raman active modes that are barely affected by anharmonicity.

\begin{table}[t]
\begin{tabular}{cccc}
\hline
\hline
P (GPa) & Mode &  $\omega_{h}$ ($\mathrm{cm}^{-1}$) & $\omega_{anh}$ ($\mathrm{cm}^{-1}$) \\
\hline
400 & $E_g$     & 1161.8  &  -     \\
~   & $B_{1g}$  & 2472.2  &  -     \\
500 & $E_g$     & 1214.6  & 1187.9 \\  
~   & $B_{1g}$  & 2745.4  & 2769.8 \\
600 & $E_g$     & 1212.4  &  -     \\
~   & $B_{1g}$  & 2988.4  &  -     \\
\hline

\hline
\end{tabular}\caption{Raman active modes of $I4_1/amd$ hydrogen 
in the 400-600 GPa pressure range. Harmonic and anharmonic frequencies are represented as 
$\omega_{h}$ and  $\omega_{anh}$ respectively.}\label{table:ramanir}
\end{table} 



\subsection{Superconductivity}\label{superconductivity}

\begin{figure}[t]
\includegraphics[width=1.0\linewidth]{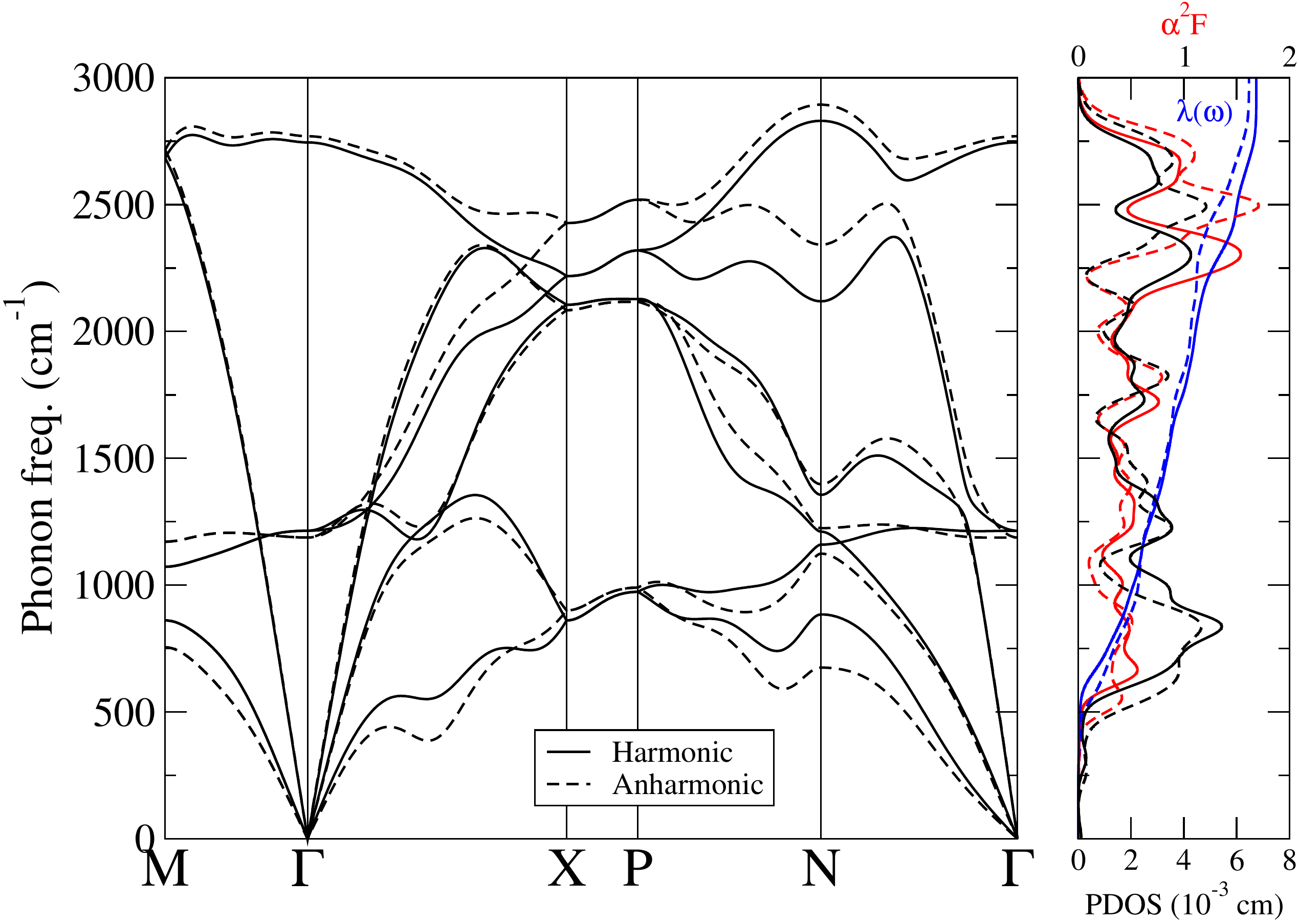}
\caption{\label{fig:phonons.anharmonic} Left: harmonic and anharmonic phonon spectra of $I4_1/amd$ 
hydrogen at 500 GPa. Right: harmonic and anharmonic phonon density Of states (PDOS), 
electron-phonon Eliashberg's function
$\alpha^2F(\omega)$ and frequency dependent electron-phonon coupling constant $\lambda(\omega)$.}
\end{figure}

In order to analyze how anharmonicity affects superconductivity, 
we have calculated Eliashberg's function $\alpha^2F(\omega)$ at 500 GPa both in the harmonic and anharmonic cases 
as described in Sec. \ref{theory}. The $\alpha^2F(\omega)$ shows a large peak at high energy
due to the large electron-phonon linewidth of high-energy optical modes (see Fig. \ref{fig:phonons.anharmonic}).
The practically linear and homogeneous increase of the integrated electron-phonon coupling constant
$\lambda(\omega)= 2 \int_{0}^{\omega} {\rm d}\omega' \alpha^2F(\omega')/\omega'$ indicates
that the contribution to the electron-phonon coupling constant is quite homogeneous over 
all the modes in the BZ. This is true both in the harmonic and in the anharmonic case.
Even if anharmonic effects have a significant impact on the phonon spectra, 
the electron-phonon coupling constant $\lambda=\lambda(\infty)$, 
which scales with the phonon frequencies as $\propto \omega^{-2}$, is practically unaltered 
by anharmonicity, as we obtain $\lambda=1.68$ and $\lambda=1.63$ in the harmonic and anharmonic cases,
 respectively.
The contribution of the low-energy transverse acoustic modes to the electron-phonon coupling 
constant
is slightly enhanced due to the anharmonic suppression of their frequencies. 
This is compensated by the hardening of the high-energy optical modes that suppresses
$\lambda$. The result is that anharmonicity slightly affects the electron-phonon coupling constant.

\begin{figure}[t]
\includegraphics[width=0.8\linewidth]{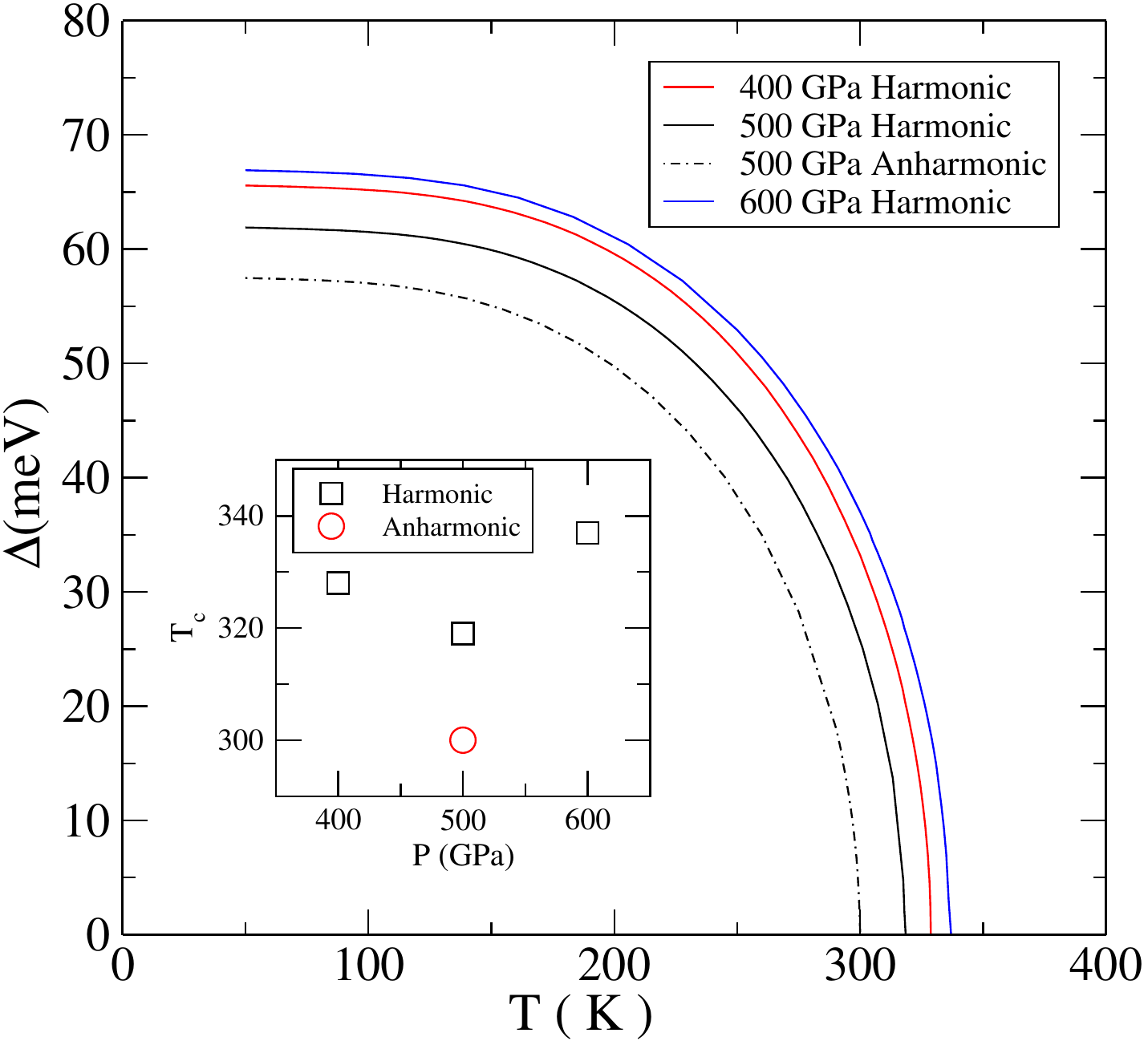}
\caption{\label{fig:supergap} Calculated first Matsubara frequency of the superconducting energy 
gap of $I4_1/amd$ hydrogen at different pressures and temperatures using a Coulomb pseudopotential $\mu^*=0.10$. 
Inset: $T_c$ vs pressure in the harmonic and anharmonic cases.
}
\end{figure}

We estimate the superconducting critical temperature by solving isotropic 
Migdal-Eliashberg equations in order to overcome the 
underestimation of $T_c$ given by McMillan's equation in the strongly interacting limit\cite{PhysRevB.12.905}. 
With a Coulomb pseudopotential of $\mu^*=0.1$ we obtain a
superconducting energy gap (first Matsubara frequency) of $\Delta\approx62$ meV and 
$\Delta\approx58$ meV, respectively, in the harmonic and anharmonic cases at $0$ K (see Fig. \ref{fig:supergap}). 
$T_c$, defined as the temperature at which the gap vanishes, is $318$ K and $300$ K respectively in the harmonic
and anharmonic approaches. The harmonic result is in reasonable agreement with previous 
results\cite{PhysRevB.84.144515}. The effect of the Coulomb pseudopotential in $T_c$ 
is weak as it happens in strongly interacting  
electron-phonon superconductors. Therefore, anharmonicity
slightly lowers the superconducting critical temperature in tetragonal $I4_1/amd$ hydrogen due to the small reduction
of $\lambda$. In Fig. \ref{fig:supergap} we also show the first Matsubara frequencies of 
the energy gap at 400 GPa and 600 GPa in the harmonic
approach. As wee can see, $T_c$ is very weakly affected by pressure in the 400-600 GPa pressure range. 
Due to the flatness of $T_c$ and the smooth change of the phonons with pressure
(Fig. \ref{fig:phonons.lindhard}) 
anharmonic effects are expected to have a similar impact at the 400 and 600 GPa.  

Considering that in all superconducting hydrides 
where strong anharmonic effects in superconductivity have been reported there are strongly softened optical modes
in the harmonic approximation\cite{PhysRevLett.114.157004,errea_inverse_isotope,PhysRevB.89.064302,
doi:10.1080/08957959.2010.520209,PhysRevB.82.104504},
even imaginary sometimes, it seems the impact of anharmonicity on $T_c$ is largely
determined by the presence of such soft hydrogen-character optical modes.
Due to the fairly uniform distribution of the electron-phonon coupling 
in the BZ here, there are no particular optical modes that soften, 
making anharmonic effects on $T_c$ weaker. 

\section{Conclusions}\label{conclusions}

In this work we have presented an exhaustive analysis of the electronic and vibrational 
properties of $I4_1/amd$ hydrogen within the 400-600 GPa pressure range. Atomic metallic hydrogen in this phase 
shows a close to free-electron-like electronic band structure, where the 
opened band gaps can be explained even without the need of electron-electron interaction. 
The huge kinetic energy of the electrons
due to the extremely high pressure plus their strong interaction with the bare nuclei makes 
the electron-electron interaction be irrelevant for the electronic 
structure. Nevertheless, 
the strong electron-proton interaction creates a big electronic localization near the atomic nuclei. 
Consequently, the RPA dramatically fails when calculating the phonons of 
atomic hydrogen. In fact, the inclusion of exchange-correlation effects in the 
calculation of the electronic response to proton motion guarantees the dynamical
stability of the structure. 

Despite anharmonicity modifies phonon frequencies up to approximately a 20\%, 
for instance, lowering the energies of the transverse acoustic modes and hardening
high-energy optical modes, it has a minor effect on superconductivity, only suppressing $T_c$
by a 6\%. This is in stark contrast to other hydrides where anharmonicity 
has a huge impact on the superconducting properties\cite{PhysRevLett.114.157004,errea_inverse_isotope,PhysRevB.89.064302,
doi:10.1080/08957959.2010.520209,PhysRevB.82.104504}, even inducing an inverse isotope effect in 
palladium hydrides\cite{errea_inverse_isotope}. This raises the interesting question whether 
anharmoncicity impacts superconductivity in hydrides simply because hydrogen is light and
vibrates far from equilibrium or for another particular reason. 
Our results suggest that determining whether anharmonicity has a strong impact
on $T_c$ cannot be related exclusively to the lightness of the ions present
in the system, but to the presence of softened optical modes. 

\section*{Appendix}\label{appendix}

Within DFT the electronic wave-functions $\ket{\psi_i}$ are calculated diagonalizing the
\begin{equation}\label{hamiltonian}
 H=T_e + V_{scf} 
\end{equation}
Hamiltonian as $H\ket{\psi_i}=\varepsilon_i\ket{\psi_i}$, where $\varepsilon_i$
is the electronic eigenvalue, $T_e$ is the electronic kinetic energy, and
the self-consistent potential is 
\begin{equation}\label{vscf}
V_{scf} = V_{e,p}+V_H+V_{xc}. 
\end{equation}
$V_{e,p}$, $V_H$, and $V_{xc}$
are, respectively, the electron-proton, Hartree, and exchange-correlation potentials.
The different band structures shown in Fig. \ref{fig:bands} are obtained neglecting 
different terms in Eq. \eqref{hamiltonian}. The DFT result, of course, retains all the
terms in $V_{scf}$; in the free-electron approximation $V_{scf} = 0$;
and in the independent electron approximation $V_{scf} = V_{e,p}$. We call the latter
approach independent electron because in this case electrons do not interact with each
other via the $V_H$ and $V_{xc}$ potentials, which depend on the electronic density $n$. 

The electronic part of the dynamical matrix $D_e$ is obtained 
Fourier transforming the electronic contribution to the force constant matrix,
which is given in terms of the electronic density and
its derivatives by 
\begin{eqnarray}
\label{second-derivatives-2}
\phi_{i,j}^{\alpha,\beta}(\mathbf{R})&=&
\int \mathrm{d}\mathbf{r} \ 
\left[ \frac{\partial n(\mathbf{r})}{\partial u_{i,\mathbf{R}}^{\alpha}} 
\right]_0 
\left[ \frac{\partial V_{e,p}(\mathbf{r}) }{\partial u_{j,0}^{\beta}} \right]_0
\nonumber \\
& + & \int \mathrm{d}\mathbf{r} \ n(\mathbf{r}) 
\left[ \frac{\partial^2 V_{e,p}(\mathbf{r})}{\partial u_{i,\mathbf{R}}^{\alpha} 
\partial u_{j,0}^{\beta}} \right]_0.
\end{eqnarray}
Here, $\mathbf{R}$ is a lattice vector and $u_{i,\mathbf{R}}^{\alpha}$ 
is the displacement in the Cartesian direction $\alpha$ of atom $i$ in the unit cell belonging to the cell
defined by $\mathbf{R}$.
The calculation of $D_e$ requires thus the knowledge of
the $\delta n$ the linear change of the electronic density with respect to to the
ionic displacements. Making use of the  
electronic density-response function $\chi(\mathbf{r},\mathbf{r}')$
the linear change of the density can be calculated as
\begin{equation}
\delta n ({\bf r}) = \int d{\bf r}' \chi(\mathbf{r},\mathbf{r}') \delta V_{e,p}(\mathbf{r}'),
\label{dn-chi}
\end{equation}
where $\delta V_{e,p}(\mathbf{r})$ represents the linear change of the electron-proton
potential.
The density response function is usually calculated by first 
estimating the noninteracting response function $\chi_0(\mathbf{r},\mathbf{r}')$,
which can be directly calculated from the eigenvalues and eigenfunctions
of the Hamiltonian in Eq. \eqref{hamiltonian} as
\begin{eqnarray}\label{chi0}
\chi^0(\mathbf{r},\mathbf{r}') &=&  
\sum_{i,j}
\frac{f_i-f_j}{\varepsilon_{i}-
\varepsilon_j}
\psi^*_{i}(\mathbf{r}) \psi_{j}(\mathbf{r}) 
\psi^*_{j}(\mathbf{r}') \psi_{i}(\mathbf{r}'),
\label{chi0-real-space}
\end{eqnarray}
where $f_i$ represents the Fermi-Dirac occupation of the
$i$-th state.
The reason for it is that the linear change of density can be 
given as
\begin{equation}
\delta n ({\bf r}) = \int d{\bf r}' \chi^0(\mathbf{r},\mathbf{r}') \delta V_{scf}(\mathbf{r}').
\label{dn-chi0}
\end{equation} 
in terms of the noninteracting response function and the linear change 
of the self-consistent potential.
Combining Eqs. \eqref{dn-chi} and \eqref{dn-chi0}, with the linear change of the
potential the following self-consistent Dyson-like equation can be obtained
for the response function:
\begin{eqnarray}\label{chi}
\chi(\mathbf{r},\mathbf{r}')&=&\chi_0(\mathbf{r},\mathbf{r}')\nonumber\\
~&+&\int d{\bf r}_1 d{\bf r}_2 \chi_0(\mathbf{r},\mathbf{r}_1) K(\mathbf{r}_1,\mathbf{r}_2)\chi(\mathbf{r}_2,\mathbf{r}'),
\end{eqnarray}
with 
\begin{equation}
K(\mathbf{r},\mathbf{r}')=\frac{1}{|\mathbf{r}-\mathbf{r}'|}+\frac{\partial^2 E_{xc}}{\partial 
n(\mathbf{r})\partial n(\mathbf{r}')}
\label{k_chi}
\end{equation}
and $E_{xc}$ the exchange-correlation energy. 
If $\delta V_{xc}(\mathbf{r})$ is neglected in Eq. \eqref{dn-chi0}, 
$K(\mathbf{r},\mathbf{r}')=\frac{1}{|\mathbf{r}-\mathbf{r}'|}$. 
This is the RPA approximation.
Neglecting different contributions for $V_{scf}$ in Eq. \eqref{vscf}
one can understand which are the important contributions
to the non-interacting $\chi_0(\mathbf{r},\mathbf{r}')$.
Similarly, neglecting different terms in $\delta V_{scf}$ in Eq. \eqref{dn-chi0}
the important contributions to $\chi(\mathbf{r},\mathbf{r}')$ can be
determined.

DFPT~\cite{RevModPhys.73.515} offers a much more efficient method to calculate $\delta n$
avoiding the cumbersome calculation of the response function, which requires a 
slowly converging sum over excited states as shown in Eq. \eqref{chi0-real-space}.  
Within DFPT 
the Sternheimer equation
\begin{equation}\label{sternheimer}
 (T_e + V_{scf} -\varepsilon_i)\ket{\delta\psi_i}=-(\delta V_{scf} -\delta\varepsilon_i)\ket{\psi_i}, 
\end{equation}
is solved self-consistently. Here, $\delta\varepsilon_i$ and $\ket{\delta\psi_i}$ are, respectively, the
linear change of the electronic eigenvalues and eigenfunctions.
Once the Sternheimer equation is solved and the $\ket{\delta\psi_i}$ states are known,
$\delta n$ can be calculated and, consequently, $D_e$~\cite{RevModPhys.73.515}.
By neglecting in Eq. \eqref{sternheimer} different terms in $V_{scf}$ and $\delta V_{scf}$ 
we are able to make different implicit approaches to $\chi_0$ and $\chi$, respectively,
as described in the previous paragraph.

\begin{figure}[t]
	\includegraphics[width=1.0\linewidth]{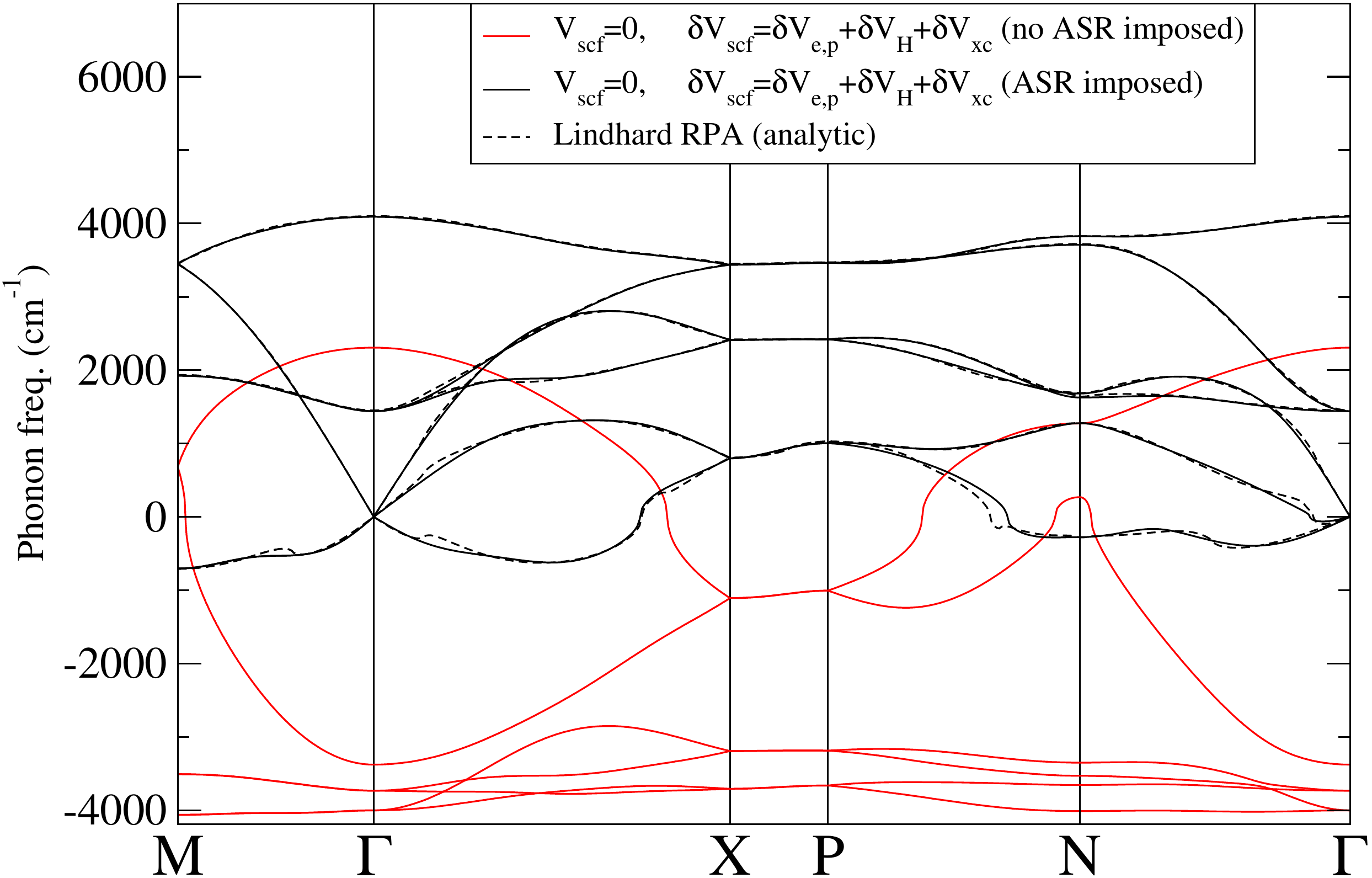}
	\caption{\label{spectraappendix1}  
Phonon spectra of $I4_1/amd$ hydrogen at 500 GPa
calculated within DFPT using $V_{scf}=0$ and $\delta V_{scf}=\delta V_{e,p}+\delta V_{H}$
with and without imposing the ASR. The results are compared to the 
Lindhard RPA spectrum calculated analytically, which is an 
equivalent calculation.}
\end{figure}

It is important to note that every time the approaches for $V_{scf}$ and $\delta V_{scf}$
are different (i.e. the neglected terms are not the same) translational invariance is broken 
down. The reason is that the ground state densities given by 
the different approximations differ. In those cases, we impose the ASR \textit{a posteriori}
by correcting the force-constants matrix in Eq. \eqref{second-derivatives-2}
as
\begin{equation}
\label{second-derivatives-3}
\tilde{\phi}_{i,i}^{\alpha,\beta}(0)  = \phi_{i,i}^{\alpha,\beta}(0) - \sum\limits_{\mathbf{R},j} \phi_{i,j}^{\alpha,\beta}(\mathbf{R}) 
\end{equation}
for every possible $i$, $\alpha$ and $\beta$. The new $\tilde{\phi}$
yield dynamical matrices that satisfy the ASR.
This way of imposing the ASR is equivalent to correcting the second addend in Eq. \eqref{second-derivatives-2}
\emph{a posteriori},
which gives a non-dispersive term to the dynamical matrix as the correction in Eq. \eqref{second-derivatives-3}.  
As an example, in Fig. \ref{spectraappendix1} we show how imposing of the ASR works in the case
$V_{scf}=0,~\delta V_{scf}=\delta V_{e,p}+\delta V_{H}$, which is equivalent to the analytic Lindhard RPA approximation. 
We see how after imposing the ASR the spectrum coincides with the analytic one. 
The small differences between the analytic spectrum and the one obtained with the
DFPT procedure are because the latter is
obtained from a Fourier interpolation from a $6\times6\times6$ \textbf{q}-grid,
while the former is calculated point by point.

\begin{figure}[t]
        \includegraphics[width=1.0\linewidth]{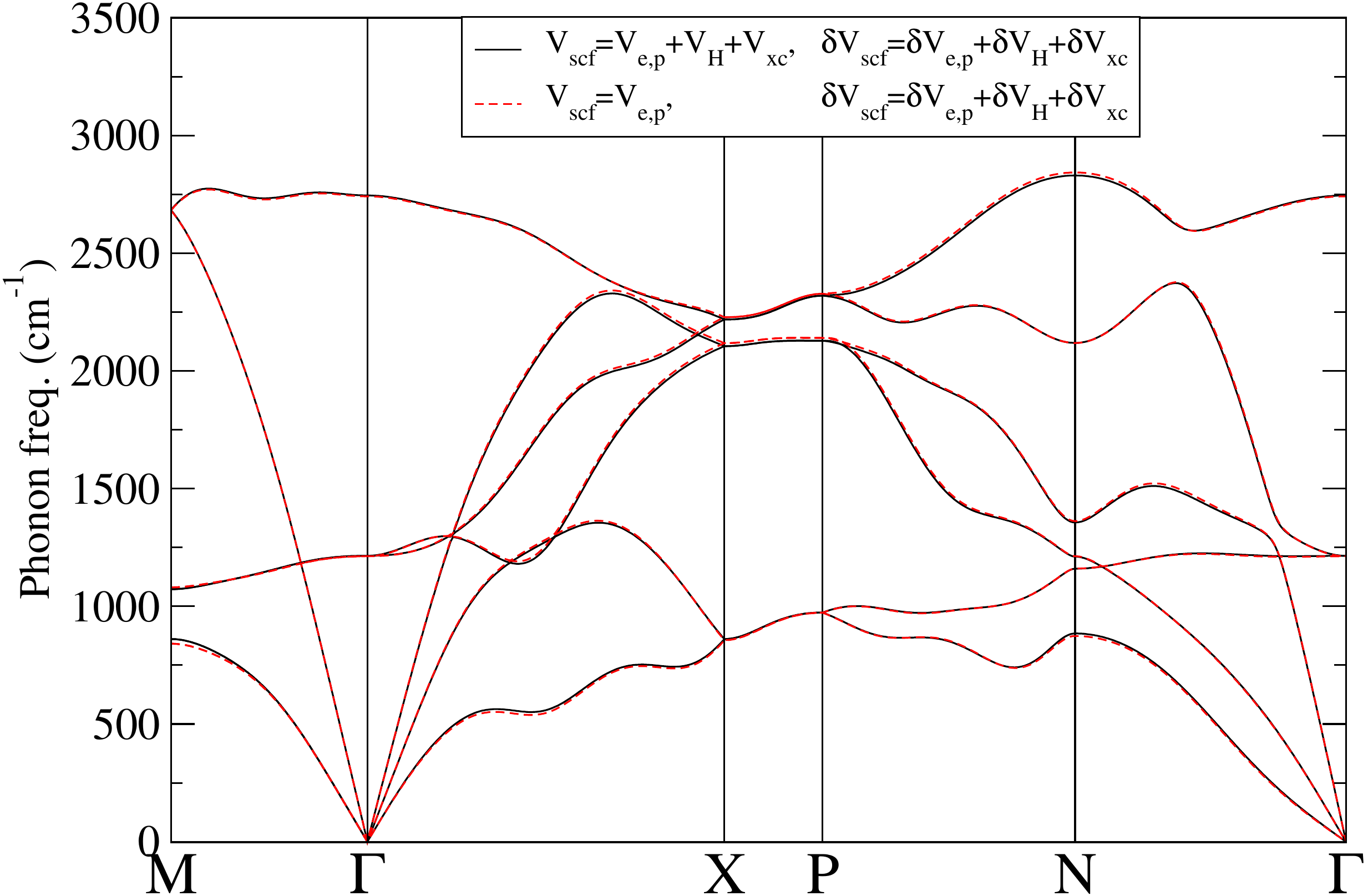}
        \caption{\label{spectraappendix2} Phonon spectra of $I4_1/amd$ hydrogen at 500 GPa
                   calculated within DFPT using two different approaches for $V_{scf}$ .
                   In the $V_{scf}=V_{e,p}$ and $\delta V_{scf}=\delta V_{e,p}+\delta V_H+\delta V_{xc}$
                   calculation the ASR is imposed \textit{a posteriori}.}
\end{figure}

The phonon-spectra presented in Fig. \ref{fig:phonons.all}
are calculated solving Eq. \eqref{sternheimer} in three different
ways: i) taking $V_{scf}=V_{e,p}$ and $\delta V_{scf}=\delta V_{e,p}+\delta V_H+\delta V_{xc}$; ii) taking
$V_{scf}=V_{e,p}$ but neglecting the linear change of the exchange correlation 
so that $\delta V_{scf}=\delta V_{e,p}+\delta V_H$; iii) neglecting any interaction of electrons in the 
self-consistent potential ($V_{scf}=0$)
but taking the full linear change of it ($\delta V_{scf}=\delta V_{e,p}+\delta V_H+\delta V_{xc}$). 
In Fig. \ref{spectraappendix2}
we show how the inclusion of the electron-electron interaction in $V_{scf}$ is irrelevant, 
as one obtains the same spectra as including it (after imposing ASR in the first case).

\section*{Acknowledgements}\label{acknowledgements}

The authors acknowledge financial support from the
Spanish Ministry of Economy and Competitiveness (FIS2013- 48286-C2-2-P), the Department of Education, Universities and Research of the Basque 
Government and the University of the Basque Country (IT756-13),  and
French Agence Nationale de la Recherche (Grant No. ANR-13-IS10-0003-01).
M.B. is also thankful to the Department
of Education, Language Policy and Culture of the Basque
Government for a predoctoral fellowship (Grant No. PRE-2014-1-477). 
Computer facilities were provided by the 
Donostia Internatinal Physics Center (DIPC).

%

\end{document}